\documentclass{aastex631}
\usepackage[version=4]{mhchem}

\begin{document}


\title[Degenerate interpretations of \ce{O3} spectral features]{Degenerate interpretations of \ce{O3} spectral features in exoplanet atmosphere observations due to stellar UV uncertainties: a 3D case study with TRAPPIST-1e}

\author[0000-0001-6067-0979]{G. J. Cooke}\thanks{E-mail: gjc53@cam.ac.uk},
\affiliation{School of Physics and Astronomy, University of Leeds, Leeds, LS2 9JT, UK\\}
\affiliation{Institute of Astronomy, University of Cambridge, UK.\\}
\author[0000-0001-6699-494X]{D. R. Marsh},
\affiliation{School of Physics and Astronomy, University of Leeds, Leeds, LS2 9JT, UK\\}
\author[0000-0001-6078-786X]{C. Walsh}
\affiliation{School of Physics and Astronomy, University of Leeds, Leeds, LS2 9JT, UK\\}
\author[0000-0002-1176-3391]{A. Youngblood}
\affiliation{NASA Goddard Space Flight Center, Solar System Exploration Division, 8800 Greenbelt Road, Greenbelt, MD 20771, USA\\}


\begin{abstract}
\noindent TRAPPIST-1e is a potentially habitable terrestrial exoplanet orbiting an ultra-cool M Dwarf star and is a key target for observations with the James Webb Space Telescope (JWST). One-dimensional photochemical modelling of terrestrial planetary atmospheres has shown the importance of the incoming stellar UV flux in modulating the concentration of chemical species, such as \ce{O3}  and \ce{H2O}. In addition, three-dimensional (3D) modelling has demonstrated anisotropy in chemical abundances due to transport in tidally locked exoplanet simulations. We use the Whole Atmosphere Community Climate Model Version 6 (WACCM6), a 3D Earth System Model, to investigate how uncertainties in the incident UV flux, combined with transport, affect observational predictions for TRAPPIST-1e (assuming an initial Earth-like atmospheric composition). We use two semi-empirical stellar spectra for TRAPPIST-1 from the literature. The UV flux ratio between them can be as large as a factor of 5000 in some wavelength bins. Consequently, the photochemically-produced total \ce{O3} columns differ by a factor of 26. Spectral features of \ce{O3} in both transmission and emission spectra vary between these simulations (e.g. differences of 19 km in transmission spectra effective altitude for \ce{O3} at 0.6 \textmu m). This leads to potential ambiguities when interpreting observations, including overlap with scenarios that assume alternative \ce{O2} concentrations. Hence, to achieve robust interpretations of terrestrial exoplanetary spectra, characterisation of the UV spectra of their host stars is critical. In the absence of such stellar measurements, atmospheric context can still be gained from other spectral features (e.g. \ce{H2O}), or by comparing direct imaging and transmission spectra in conjunction.
\end{abstract}



\section{Introduction}
\label{Introduction section}

On account of their frequency and relative ease of characterisation, planetary systems orbiting M dwarf stars are prime targets in the search for potentially habitable exoplanets. TRAPPIST-1 is an M8.5V star, orbited by seven terrestrial exoplanets, and each of them could be tidally locked to their host star \citep{2017Natur.542..456G}. TRAPPIST-1e is of particular interest because with current knowledge it is more likely than the close-in planets to have retained its atmosphere due to lower predicted escape rates \citep{2018PNAS..115..260D}, and is probably more able to sustain surface liquid water when compared to the outer planets which receive less stellar irradiation \citep{2017Natur.542..456G, 2017ApJ...839L...1W}. To begin to characterise the TRAPPIST-1 exoplanets and determine the composition of their atmospheres, at the time of writing, several observational programs with JWST are scheduled, including the observation of four transits of TRAPPIST-1e in 2023 (see program 1331)\footnote{\href{https://www.stsci.edu/jwst/science-execution/program-information.html?id=1331}{https://www.stsci.edu/jwst/science-execution/program-information.html?id=1331}, accessed Wed April 12 2023}.

In general, the detection of molecular oxygen (\ce{O2}) on an exoplanet is of profound interest because of its importance for life on Earth \citep{2003AsBio...3..689S,2018AsBio..18..630M}. Ozone (\ce{O3}) is produced in exoplanetary atmospheres by UV radiation which is able to dissociate \ce{O2}, and it has been calculated that in some situations the detection of \ce{O3} is easier to achieve than a detection of \ce{O2}; for example, for low \ce{O2} concentrations like those potentially present during Earth's Proterozoic eon, but where \ce{O3} concentrations are still detectable \citep{2022A&A...665A.156K}. Thus, in such cases, it has been proposed that the detection of \ce{O3} may be used as a proxy to confirm the presence of \ce{O2} \citep{1993A&A...277..309L, 2003AsBio...3..689S, 2018AsBio..18..630M, 2021ExA...tmp..118Q}.

One-dimensional (1D) photochemical modelling has demonstrated that planetary atmospheric composition (including \ce{O3} and \ce{H2O}) is influenced by the strength and shape of the incoming ultraviolet (UV) radiation from the host star \citep[see][and references therein]{2014P&SS...98...66G, 2013AsBio..13..251R, 2022A&A...665A.156K, 2018AsBio..18..630M}. For example, \cite{2022ApJ...927...90T} used MUSCLES Treasury survey M-dwarf spectra combined with UV spectra reconstructions as stellar spectra input to Atmos (a coupled 1D photochemistry and climate model), and similarly demonstrated that UV irradiation can modulate hydrocarbon haze concentrations. Because the atmospheric composition with respect to altitude affects molecular detectability in remote sensing, the link between \ce{O3} abundance and \ce{O2} abundance will be difficult to ascertain because it depends on several parameters, including the catalytic cycles that remove \ce{O3} (e.g. HO\textsubscript{x}, NO\textsubscript{x}, ClO\textsubscript{x} and BrO\textsubscript{x} chemical families), and atmospheric pressure. A well-characterised spectrum of the host star is required for confident modelling of planetary climate \citep{2020A&A...639A..99E}, atmospheric chemistry \citep{2022A&A...665A.156K}, and atmospheric escape \citep{2018PNAS..115..260D}. However, the host star's spectral energy distribution may not be known to high precision when analysing and interpreting exoplanet observations. 

The UV flux from TRAPPIST-1 remains uncertain because of the intrinsic faintness of the star \cite[$V=18.798$~mag;][]{2006AJ....132.1234C}. \citet{2019ApJ...871..235P}, henceforth known as P19, modelled the spectrum of TRAPPIST-1. To do this, they used the PHOENIX stellar atmospheric code \citep{2007A&A...468..255B,1993JQSRT..50..301H,2006A&A...451..273H}, and added a treatment of the chromosphere to produce synthetic stellar spectra of cool dwarf stars, including TRAPPIST-1, whose ultraviolet light have negligible flux contribution from the photosphere. More recently, \citet{2021ApJ...911...18W}, hereafter known as W21, used new observations of TRAPPIST-1 to create a semiempirical spectrum for use in atmospheric modelling simulations. Whilst neither spectrum wholly represents the true stellar irradiation environment of the TRAPPIST-1 planets, the W21 spectrum is in significantly better agreement with available observations of TRAPPIST-1.

In addition to the incoming stellar spectrum, the 3D transport and chemistry of the exoplanet's atmosphere is important for understanding the distribution and abundance of chemical species. \cite{2019ApJ...886...16C} investigated exoplanets orbiting at the inner edge of the habitable zone using a 3D chemistry climate model (WACCM4), showing how different assumed UV spectra can influence atmospheric mixing ratios of species such as \ce{H2O}, \ce{O3}, and \ce{H}. Additionally, 3D simulation studies have demonstrated the influence of UV radiation on 3D transport \citep{2019ApJ...886...16C}, stratospheric temperature \citep{2015P&SS..111...62G, 2019ApJ...886...16C}, and the distribution and abundance of chemical species \citep{2018ApJ...868L...6C, 2019ApJ...886...16C, 2021NatAs...5..298C, 2022MNRAS.517.2383B, 2022arXiv221013257R}. Tidally locked terrestrial exoplanets modelled in 3D exhibit atmospheric jets that transport heat and chemical constituents to the night side \citep{2011ApJ...738...71S,2020A&A...639A..99E, 2020MNRAS.492.1691Y}. \citet{2016EP&S...68...96P}, \citet{2018ApJ...868L...6C} and \citet{2020MNRAS.492.1691Y} found that \ce{O3}, which is photochemically generated on the dayside, can be transported to the night side, where its lifetime increases due to the lack of UV irradiation and a reduction in catalytic cycle destruction. This body of previous work motivates the need to use 3D models when investigating the climate and chemistry of specific exoplanets, in particular, with respect to their molecular observability linked to the oxygenation state of the atmosphere.

In this study we simulate the atmosphere of TRAPPIST-1e using the WACCM6 Earth System Model and the stellar spectra from P19 (\footnote{\href{http://archive.stsci.edu/hlsp/hazmat}{doi:10.17909/t9-j6bz-5g89}}model 1A, version 1) and W21 (\footnote{\href{https://zenodo.org/record/4556130\#.Y_9hOOb7RmM}{https://zenodo.org/record/4556130\#.Y\textunderscore 9hOOb7RmM}} version 7) which differ in UV flux in some wavelength bins by up to a factor of 5000. Our aim is to quantify the effects of such uncertainties in the strength of UV from the host star on the climate and composition of the atmosphere. This is the first time a 3D global climate model has been used to simultaneously assess the influence of uncertain UV spectra and transport on the climate and chemistry of TRAPPIST-1e. Additionally, we simulate possible future observations (transmission and emission spectra) of TRAPPIST-1e using the outputs from the WACCM6 simulations. We discuss the implications of uncertainties in the stellar UV spectrum in the interpretation of future observations of terrestrial exoplanets.

\newpage

\section{Methods}
\label{Methods section}

\subsection{UV spectra input}
\label{UV input spectra}

This work employs two different assumed stellar spectra as input to the simulations. P19 \citep{2019ApJ...871..235P} generated three different models of TRAPPIST-1's spectrum (Model 1A, 2A, and 2B). Model 1A was created such that the emission was benchmarked to the Ly-$\alpha$ reconstruction that was presented in \cite{2017A&A...599L...3B}, who used the Hubble Space Telescope (HST) Space Telescope Imaging Spectrograph instrument to observe the TRAPPIST-1 Ly-$\alpha$  line. Using an alternative approach to construct Model 2A and Model 2B, P19 aligned the stellar emission to be within with the range of distance-corrected Galaxy Evolution Explorer NUV photometry of stars with a spectral type akin to TRAPPIST-1, whilst remaining compatible with FUV upper limits. The EUV estimates were extracted from empirical scaling relationships based on X-ray and
Ly-$\alpha$ emission.

W21 \citep{2021ApJ...911...18W} used new HST (1100 -- 5500 \AA) and XMM-Newton (10 -- 50 \AA) observations of TRAPPIST-1 from the Mega-MUSCLES (Measurements of the Ultraviolet Spectral Characteristics of Low-Mass Exoplanetary Systems) Treasury Survey \citep{2019ApJ...871L..26F, 2021ApJ...911...18W} to construct a 5 \AA \ -- 100 \textmu m spectrum of the star. Four models were used to fill in gaps in wavelength coverage, including a PHOENIX model for wavelengths $>5500$ \AA\ ($>0.55$ \textmu m). Because of TRAPPIST-1's relatively low luminosity, W21 substituted the noisy 1100 -- 4200 \AA\ HST spectrum with a semi-empirical, noiseless spectrum that reproduced the measured flux of detected UV emission lines and agreed with the upper limits on the stellar continuum established by the HST spectra.

W21 found that whilst the P19 Model 1A shows good agreement with the C II and Ca II stellar lines, several other lines are inconsistent with measured fluxes, exhibiting a discrepancy $\sim 10$ times more than the W21 upper limits permit. In the WACCM6 simulations, we decided to use both the P19 and W21 spectra in order to illustrate important differences that may occur in instances where the exoplanet's photochemical environment is highly uncertain.

\subsection{WACCM6}
\label{The climate model WACCM6 section}

TRAPPIST-1 is an ultra cool M dwarf at a distance of 12.4 pc. Its stellar properties are summarised in Table~\ref{Properties table}. We assume that TRAPPIST-1e receives $900~\textrm{W\,m}^{-2}$ of irradiation (0.66\ $S_\oplus$, where $S_\oplus$ is the total insolation received by the Earth). This is consistent with the value used in the TRAPPIST-1 Habitable Atmosphere Intercomparison project \citep[THAI;][]{2020GMD....13..707F,2022PSJ.....3..211T}, although note that the latest data available in the NASA Exoplanet Archive \citep{2013PASP..125..989A} lists a value of $0.646\pm0.025\ S_\oplus$ \citep{2021PSJ.....2....1A}.

We use the Earth System Model WACCM6 to simulate the climate of TRAPPIST-1e with the properties indicated in Table~\ref{Properties table}, using given values from  from \cite{2018MNRAS.475.3577D} and \cite{2018A&A...613A..68G}. Note that the simulations were started in 2020, before \cite{2021PSJ.....2....1A} published their work. WACCM6 is a specific configuration of the Community Earth System Model version 2 (CESM2). The model release we use in this work is CESM2.1.3\footnote{\href{http://www.cesm.ucar.edu/models/cesm2/}{http://www.cesm.ucar.edu/models/cesm2/}}. We use a pre-industrial atmosphere (approximating the atmosphere of 1850 in terms of pollutants and greenhouse gas mixing ratios) with the modern ocean and land configuration, a horizontal resolution of $1.875^\circ$ by $2.5^\circ$ (latitude by longitude), and 70 vertical atmospheric levels. The ocean and atmosphere are fully interactive, meaning that they respond to physical perturbations such as temperature, or in the case of the atmosphere, chemical perturbations. Because TRAPPIST-1e is suspected to be tidally locked \citep{2017Natur.542..456G}, we lock the substellar point by fixing the solar zenith angle in each grid cell, and we set the exoplanet’s obliquity and orbital eccentricity to zero \citep[note that the eccentricity may be non-zero, albeit $<0.01$;][]{2017NatAs...1E.129L}. The substellar point is placed the Pacific Ocean at $180^\circ$ longitude and $0^\circ$ latitude. The chemical mechanism, which is described in \cite{2020JAMES..1201882E}, has 98 chemical species and 298 chemical reactions (including photochemical reactions). Absorption by \ce{CO2} and \ce{H2O} in the Schumann-Runge bands is included \citep{2023RSOS...1030056J}. The full details of the model set-up, alongside simulation scripts, are available via GitHub\footnote{\href{https://github.com/exo-cesm/CESM2.1.3/tree/main/Tidally_locked_exoplanets}{https://github.com/exo-cesm/CESM2.1.3/tree/main/Tidally\textunderscore locked\textunderscore exoplanets}}.

\begin{table}[t!]
\caption{Orbital and planetary parameters used in this study, for the exoplanet TRAPPIST-1e and its star (TRAPPIST-1). The parameters are from \citet{2018MNRAS.475.3577D}, \citet{2018A&A...613A..68G}, and \citet{2021PSJ.....2....1A}. The mass and radius are chosen to be consistent with the THAI project \citep{2020GMD....13..707F,2022PSJ.....3..212S, 2022PSJ.....3..211T}. $\odot$ and $\oplus$ subscripts denote values relative to the Sun and Earth, respectively.}
\centering
\begin{tabular}{@{}lll@{}}
\toprule
Parameter &  Value \\ \hline
Stellar luminosity [$L_\odot$] & 0.000553 \\
Stellar effective temperature [K] & 2566 \\
Stellar radius [$R_\odot$] &  0.119  \\
Stellar mass [$M_\odot$] & 0.089 \\
Stellar metallicity &  0.04 \\
Planetary orbital period [days] &  6.099 \\
Planetary radius [$R_\oplus$] &  0.91 \\
Planetary mass [$M_\oplus$] & 0.772 \\
Planetary surface gravity [m s$^{-2}$] & 9.1454 \\\hline
\label{Properties table}
\end{tabular}
\end{table}

\begin{figure*}[t!]
	\centering
	\includegraphics[width=1\textwidth]{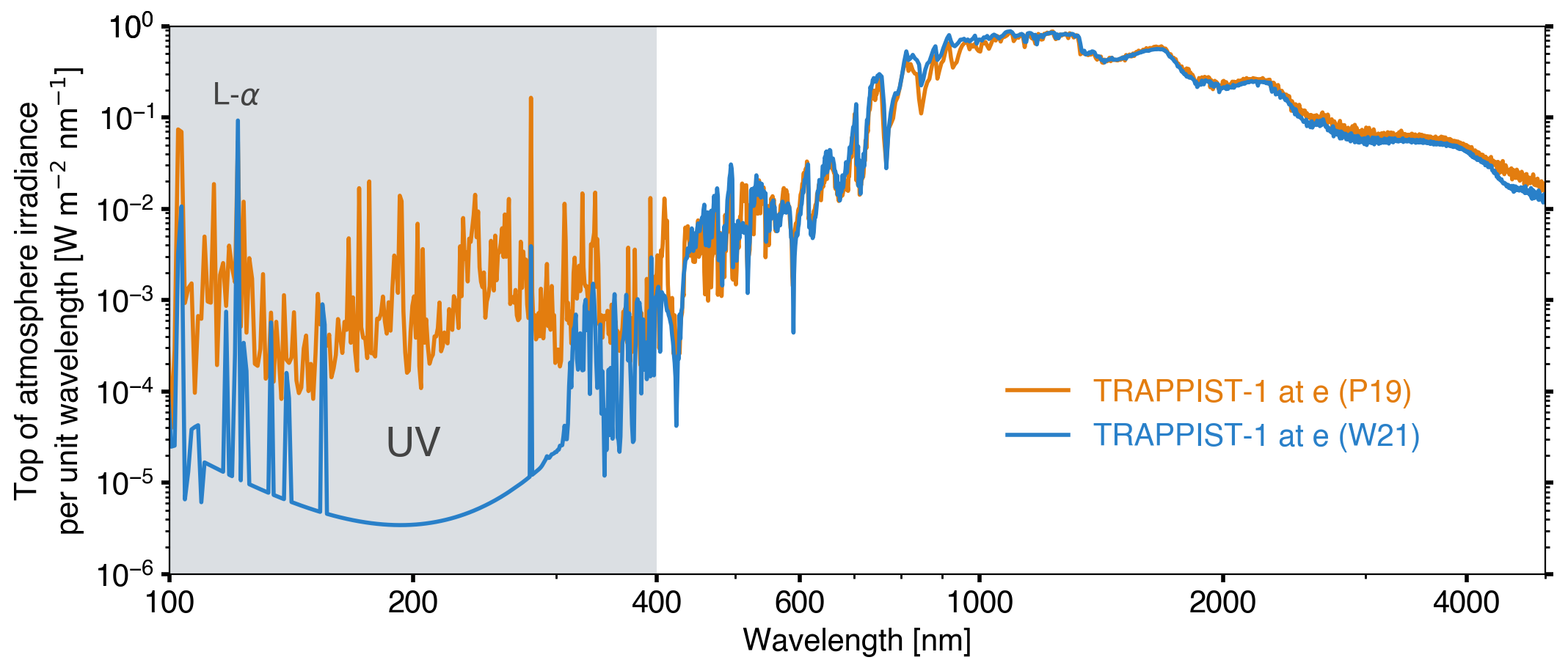}
    \caption{The top of atmosphere irradiance per unit wavelength as a function of wavelength for the TRAPPIST-1 semi-empirical stellar spectra from \citet[][orange]{2019ApJ...871..235P} and \citet[][light blue]{2021ApJ...911...18W}, each of which are scaled to the total integrated irradiance received by TRAPPIST-1e ($900~\mathrm{W\,m}^{-2}$). These spectra were rebinned in order to match the wavelength grid used in WACCM6 simulations. The grey shaded region indicates UV wavelengths (100 -- 400 nm).} 
    \label{Stellar spectra figure}
\end{figure*}

We scale the P19 and W21 spectra to the irradiance received by TRAPPIST-1e ($900~\textrm{W\,m}^{-2}$), rebinning them to match the wavelength grid required for WACCM6 simulations. The resulting spectra are shown in Fig.~\ref{Stellar spectra figure}. 
For the wavelength regions over which \ce{O2} and \ce{O3} photolyse, the integrated flux is listed in Table~\ref{Flux table}.

We present ten simulations with WACCM6, five with the P19 spectrum, and five with the W21 spectrum -- see Table~\ref{Run table} for a summary. The initial conditions for the five simulations with each spectrum consist of one with the standard initial pre-industrial Earth composition (PI), then three lower \ce{O2} simulations with a composition with 10, 100, and 1000 times less \ce{O2} (10\% PAL, 1\% PAL, and 0.1\% PAL), and one in which the planet is not tidally locked (noTL). These simulations allow us to assess the influence of different strengths of incoming UV spectra, the influence of tidal locking, and the effects of reducing \ce{O2} in order to quantify any degeneracies between \ce{O2}, \ce{O3}, and incident UV light. Each simulation is run for over 250 model Earth years, and the last ten years of the simulation are used for time-averaged results (Figs \ref{Ox production figure} -- \ref{O2 vs O3 column figure} and Figs \ref{Surface temperatures figure} -- \ref{Cloud fraction figure}). The WACCM6 simulations were run on 120 cores at a model cost of 1,332 pe-hours per simulated year. A 250 year simulation therefore costs 333,000 pe-hours.

\begin{table*}[t!]
\centering
\caption{For the two different spectra used in the simulations, the integrated flux (in units of W m$^{-2}$) is given for each spectrum for the total flux and six different wavelength bands: Schumman-Runge continuum (S-RC), Schumman-Runge bands (S-RB), Herzberg continuum (HC), Hartley band (HaB), Huggins band (HuB), and the Chappuis band (CB). For reference, the Earth receives 1360 $\textrm{Wm}^{-2}$ of irradiation. Photons in the Schumman-Runge continuum, Schumann-Runge bands, and Herzberg continuum are able to photolyse \ce{O2}. Photons in the Hartley band, Higgins band, and Chappuis band, are able to photolyse \ce{O3}.}
\label{Flux table}
\begin{tabular}{@{}cccccccc@{}}
\toprule
Spectrum & Total & S-RC & S-RB & HC & HaB & HuB & CB \\
 & 10.5--99975 nm & 130--175 nm &  176-192 nm & 200-240 nm & 200-310 nm&  310--340 nm & 400-650 nm\\\hline
Sun & 1361 & 0.0092 & 0.0373 & 1.2997 & 19.8916 & 23.4556 & 453.7313 \\
TP-1 P19 & 900 & 0.0382 & 0.0239 & 0.0841 & 0.3953 & 0.0829 & 2.1355 \\
TP-1 W21 & 900 & 0.0025 & 0.0001 & 0.0002 & 0.0050 & 0.0133 & 2.3045 \\
Flux ratio $\frac{\textrm{P19}}{\textrm{W21}}$ & 1.00 & 15.54 & 451.81 & 528.85 & 79.59 & 6.21 & 0.93 \\
\hline
\end{tabular}
\end{table*}

\begin{table*}[t!]
\centering
\caption{Ten simulations have been performed, each with an obliquity of $0^\circ$ and a circular orbit, using two different spectra for TRAPPIST-1 from P19 and W21 (see text for UV spectra input details). Eight simulations were set up in a tidally locked configuration with a 6.1 day rotation rate. The ``PI" simulations have an initial pre-industrial Earth composition and are the same apart from using the P19 and W21 spectra. The ``noTL" simulations have an initial pre-industrial Earth composition, are not tidally locked, and have a diurnal cycle (rotational period of 1 day). For the lower \ce{O2} scenarios, the volume mixing ratio of atmospheric \ce{O2} is reduced by a factor of 10, 100, and 1000 (the ``10\% PAL",``1\% PAL", and ``0.1\% PAL" simulations, respectively). PAL means present atmospheric level, where the present atmospheric level of oxygen is a volume mixing ratio of 0.21.}
\begin{tabular}{@{}cccc@{}}
\toprule
Simulation & Spectrum & \ce{O2} mixing ratio [PAL] & Orbital parameters \\ \hline
P19 PI & P19 & 1 & Tidally locked, 6.1 day rotational period \\
P19 10\% PAL & P19 & 0.1 & Tidally locked, 6.1 day rotational period \\
P19 1\% PAL & P19 & 0.01 & Tidally locked, 6.1 day rotational period \\
P19 0.1\% PAL & P19 & 0.001 & Tidally locked, 6.1 day rotational period \\
P19 noTL & P19 & 1 & Not tidally locked, 1 day rotational period \\
W21 PI & W21 & 1 & Tidally locked, 6.1 day rotational period \\ 
W21 10\% PAL & W21 & 0.1 & Tidally locked, 6.1 day rotational period \\
W21 1\% PAL & W21 & 0.01 & Tidally locked, 6.1 day rotational period \\
W21 0.1\% PAL & W21 & 0.001 & Tidally locked, 6.1 day rotational period \\
W21 noTL & W21 & 1 & Not tidally locked, 1 day rotational period \\\hline
\label{Run table}
\end{tabular}
\end{table*}

\subsection{Planetary Spectrum Generator}
\label{The climate model WACCM6 section}

We use the Planetary Spectrum Generator \citep[PSG;][]{2018JQSRT.217...86V} GlobES\footnote{ \href{https://psg.gsfc.nasa.gov/apps/globes.php}{https://psg.gsfc.nasa.gov/apps/globes.php}} 3D mapping tool to compute transmission and emission spectra from the WACCM6 atmospheric simulations, using instantaneous data (a `snapshot' at a model time step rather than an average over model time steps) for each produced spectrum. The WACCM6 instantaneous data are rebinned to a resolution of $10^{\circ}$ in longitude only and we keep the same latitudinal grid resolution ($1.875^{\circ}$). This is done to reduce the data size so it is compatible with GlobES in PSG, and is the same approach as used in \citet{2023MNRAS.518..206C}. The same data at the same time step is used for both transmission spectra and emission spectra in this paper. PSG ingests the data and integrates across the whole observable disk to produce a reflection or emission spectrum. It uses the grid cells at the terminator to produce a transmission spectrum. Because model grid cells either side of the terminator would contribute to the opacity of the atmosphere, this is not the most realistic way to represent the atmospheric geometry during a planetary transit \citep{2019A&A...623A.161C}. However, it is adequate for the relatively low temperature contrasts between the day and night sides for the TRAPPIST-1e atmospheres simulated here.

The molecules we use for the computation of the transmission and emission spectra are \ce{N2}, \ce{O2}, \ce{CO2}, \ce{H2O}, \ce{O3}, \ce{CH4}, and \ce{N2O} which is a list that includes possible biosignatures and indicators of habitability \citep{2017ARA&A..55..433K}. The default HITRAN opacity data \citep{2017JQSRT.203....3G} is used for each molecule, as well as all available collision-induced absorption coefficients (e.g. \ce{O2-O2}, \ce{N2-N2}, and \ce{O2-N2}). The radiative transfer model used in PSG is the Planetary and Universal Model of Atmospheric Scattering (PUMAS) model. The correlated-k method is used by PUMAS for the spectral resolving powers used in this paper \citep{2019A&A...623A.161C}. If high resolution is required, it would instead use the line-by-line method. Scattering effects are included, as are ice clouds and water clouds. The effective radius of the cloud particles is assumed to be 5 \textmu m for water and 100 \textmu m for ice clouds. Scripts and data to generate the PSG files are provided with the data associated with this article.

\section{Results}
\label{Results section}

\subsection{Atmospheric chemistry}

\begin{figure*}[t!]
	\centering
	\includegraphics[width=\textwidth]{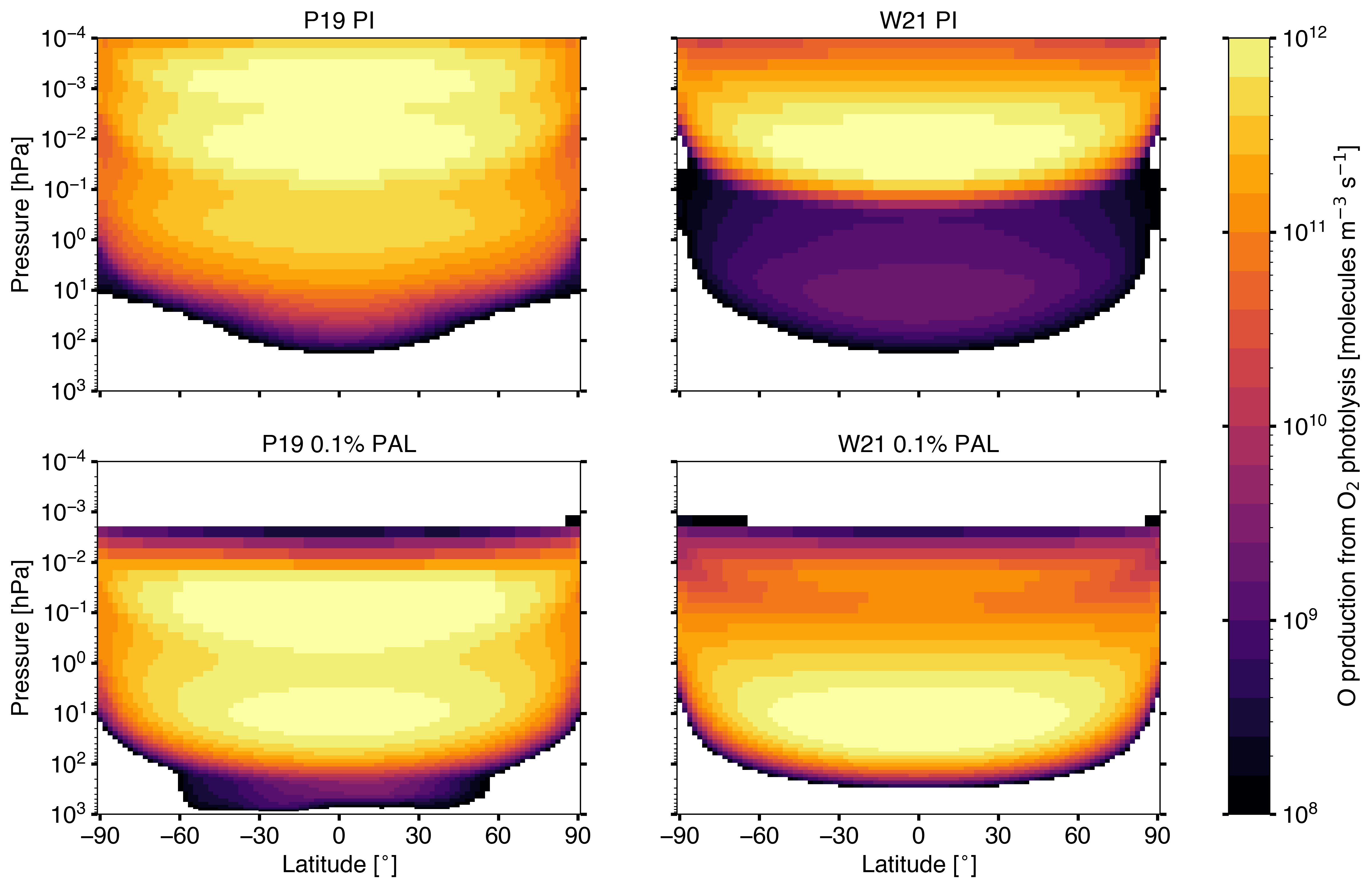}
    \caption{The P19 (left column) and W21 (right column) irradiation scenarios are compared to each other for the PI (top row) and 0.1\% PAL (bottom row) cases. The zonal mean (averaged over longitude) of the production of O from \ce{O2} photolysis is displayed, with latitude in $^\circ$ on the horizontal axis and pressure in hPa on the vertical axis. \ce{O2} photolysis takes place at wavelengths less than 242 nm. The white regions show where the production has dropped below $10^8$ molecules m\textsuperscript{-3} s\textsuperscript{-1}.} 
    \label{Ox production figure}
\end{figure*}

Atmospheric \ce{O3} is created and destroyed via the Chapman cycle \citep{chapman1930xxxv}, which is initiated by ultraviolet light (of wavelength $\lambda$) dissociating an \ce{O2} molecule:

\begin{align}
     \ce{O2 +} h\nu\ (175.9\ \textrm{nm} < \lambda < 242.4\ \textrm{nm}) &\ce{-> O + O},\\
      \ce{O2 +} h\nu\ (\lambda < 175.9\ \textrm{nm} ) &\ce{-> O(^1D) + O(^3P)},\\
    \ce{O + O2 + M &-> O3 + M}\label{ozone prod},\\ 
    \ce{O + O3 &-> 2 O2}\label{ozone loss},\\
    \ce{O3 +} h\nu\ (\lambda \geq 320\ \textrm{nm}) &\ce{-> O2(^3\Sigma^-_g) + O(^3P)},\\
    \ce{O3 +} h\nu\ (\lambda \leq 320\ \textrm{nm}) &\ce{-> O2(^1\Delta_g) + O(^1D)}.
\end{align}

The latter two photolysis reactions do not contribute to \ce{O3} destruction because the O produced rapidly recombines with \ce{O2} to produce \ce{O3} via reaction \ref{ozone prod}. However, the reaction between O and \ce{O3} (reaction \ref{ozone loss}) does lead to a loss of \ce{O3}, and it can be sped up through catalytic agents, which are denoted below as X. An example of a catalytic cycle is shown here:

\begin{equation}
    \begin{split}
        \ce{X + O3 &-> XO + O2},\\
        \ce{XO + O &-> X + O2}, \\
        \textrm{Overall:}\ \ce{O3 + O &-> 2 O2}.
    \end{split}
\end{equation}

Catalytic agents may be NO, H, OH, Cl, or Br \citep{2005ama..book.....B}. These species can be produced through other photolysis reactions (e.g., \ce{H2O} photolysis producing H and OH).

\begin{figure*}[b!]
	\centering
	\includegraphics[width=\textwidth]{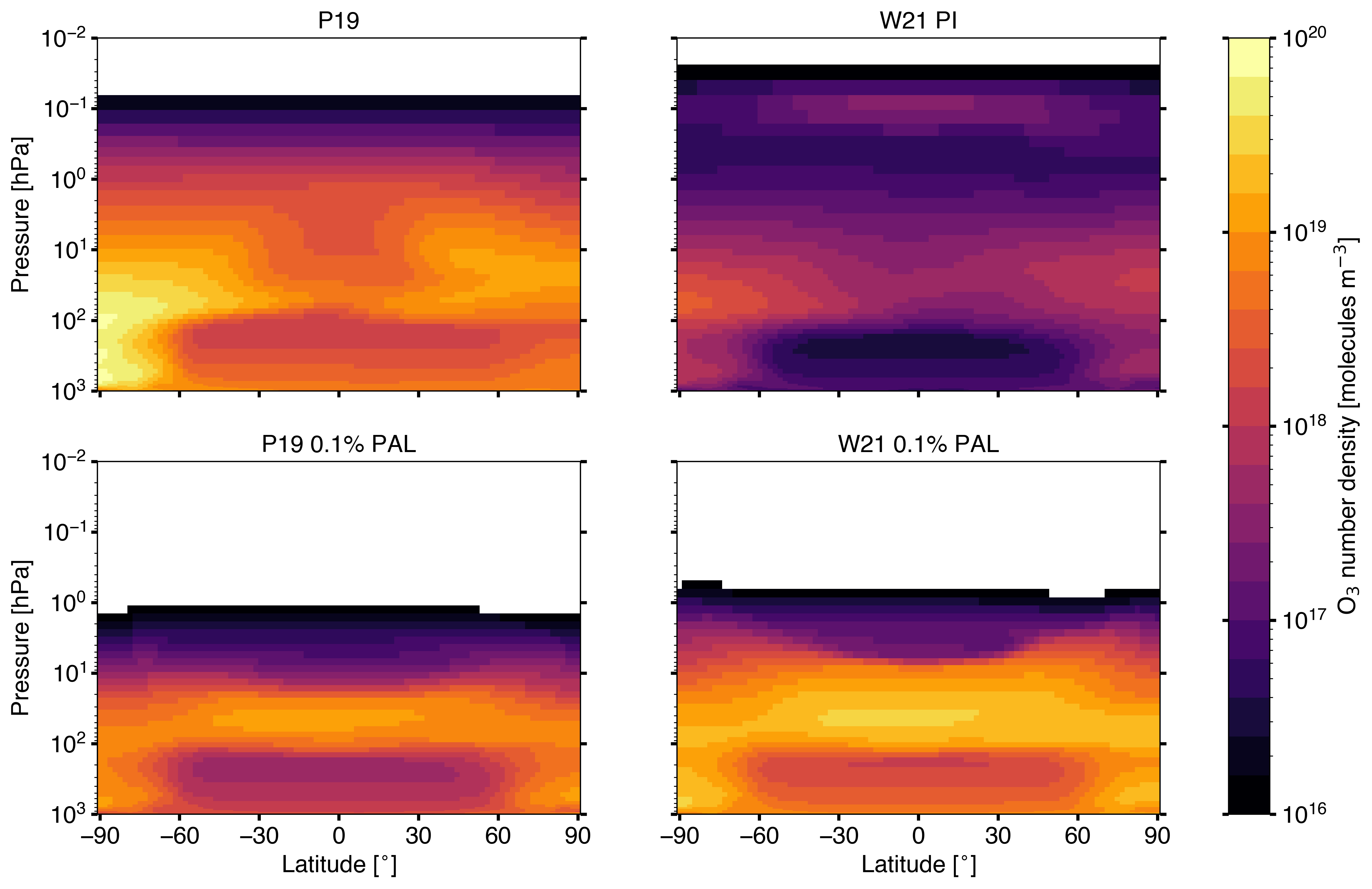}
    \caption{The P19 (left column) and W21 (right column) irradiation scenarios are compared to each other for the PI (top row) and 0.1\% PAL (bottom row) cases. The zonal mean (average over longitude) of the \ce{O3} number density is shown, with latitude in $^\circ$ on the horizontal axis and pressure in hPa on the vertical axis. The white regions show where the number density has dropped below $10^{16}$ molecules m\textsuperscript{-3}.} 
    \label{O3 density figure}
\end{figure*}

Fig.~\ref{Ox production figure} shows the zonal mean of O production from \ce{O2} photolysis in the P19 PI, P19 0.1\% PAL, W21 PI, and W21 0.1\% PAL simulations. In the two PI simulations, \ce{O2} can be photolysed lower down in the P19 case because there is more incoming UV radiation than in the W21 case. When \ce{O2} is reduced to 0.1\% PAL, it can be seen that the peak of \ce{O2} photolysis moves to higher pressures (downward in altitude). Where O is produced impacts the number density of \ce{O3}. Fig.~\ref{O3 density figure} shows the zonal mean of the \ce{O3} number density for the same set of simulations as in Fig.~\ref{Ox production figure}. Reducing \ce{O2} in the P19 cases causes a drop in \ce{O3} number density at pressures above 1 hPa, whilst the opposite occurs in the W21 cases. This is because of the pressure dependency on the reaction that produces \ce{O3} (reaction \ref{ozone prod}).

\begin{figure*}[t!]
	\centering
	\includegraphics[width=\textwidth]{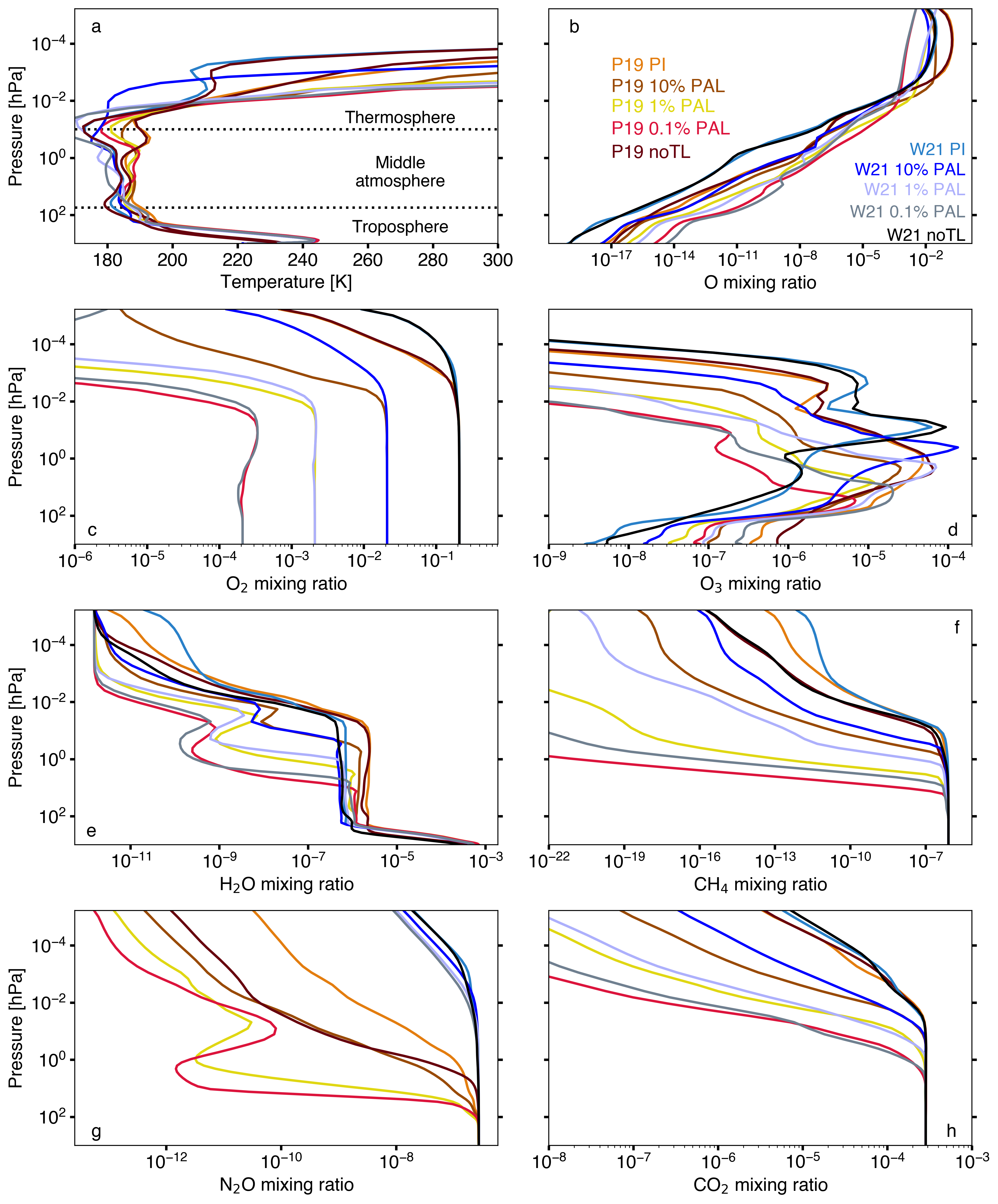}
    \caption{The global mean temperature (a) is plotted against pressure for the P19 PI (orange), P19 noTL (dark red), P19 10\% PAL (brown), P19 1\% PAL (yellow), and P19 0.1\% PAL (red) simulations, and the W21 PI (light blue), W21 noTL (black), W21 10\% PAL (blue), W21 1\% PAL (lilac), and W21 0.1\% PAL (grey) simulations. The globally averaged mixing ratios for \ce{O} (b), \ce{O2} (c), \ce{O3} (d), \ce{H2O} (e), \ce{CH4} (f), \ce{N2O} (g), and \ce{CO2} (h), are also shown.} 
    \label{Chemistry figure}
\end{figure*}

In Fig.~\ref{Chemistry figure} we show the global mean vertical profiles for temperature, and the mixing ratio of \ce{O}, \ce{O2}, \ce{O3}, \ce{H2O}, \ce{CH4}, \ce{N2O}, and \ce{CO2}. On global average, the temperature profile shows deviations of up to 9 K in the troposphere and up to 19 K below the thermosphere between the W21 PI and P19 PI simulations. The W21 PI middle atmosphere (between the troposphere and thermosphere) is colder than the P19 PI middle atmosphere due to reduced \ce{O3} heating because of the lower \ce{O3} concentration. The P19 noTL simulation has a lower temperature in the troposphere by up to 23 K and in the middle atmosphere by up to 6 K, resulting in lower concentrations of \ce{H2O} compared to the P19 PI simulation. Generally, between 200 and 10 hPa, the relative \ce{H2O} number density in each simulation correlates with the relative temperature profile in each simulation (i.e. a  lower temperature results in less \ce{H2O}).

The \ce{O3} mixing ratio profile in Fig.~\ref{Chemistry figure} shows large deviations between the P19 PI (orange) and W21 PI (light blue) simulations, with a difference of a factor of 116 in the mixing ratio at the surface. The \ce{O3} number density peaks at $2.0\times10^{19}$ molecules m$^{-3}$ at 50 hPa in the P19 PI case and at $6.3\times10^{17}$ molecules m$^{-3}$ at 50 hPa in the W21 PI case. Despite the amount of \ce{O3} present in the P19 PI simulated atmosphere, the temperature inversion in the middle atmosphere is $<8$ K because the UV intensity is not sufficient to provide enough \ce{O3} UV heating to create a similar temperature inversion to Earth's stratosphere ($\approx60$ K change between the tropopause and stratopause). The P19 noTL simulation (dark red) shows that tidal locking increases the amount of \ce{O3} in the middle atmosphere, whilst reducing the \ce{O3} concentration in the troposphere. The non-tidally locked cases are colder in the lower atmosphere, and warmer in the middle atmosphere. This reduces the rate of catalytic destruction reactions and allows \ce{O3} formation to occur faster. 

\begin{figure*}[t!]
	\centering
	\includegraphics[width=1\textwidth]{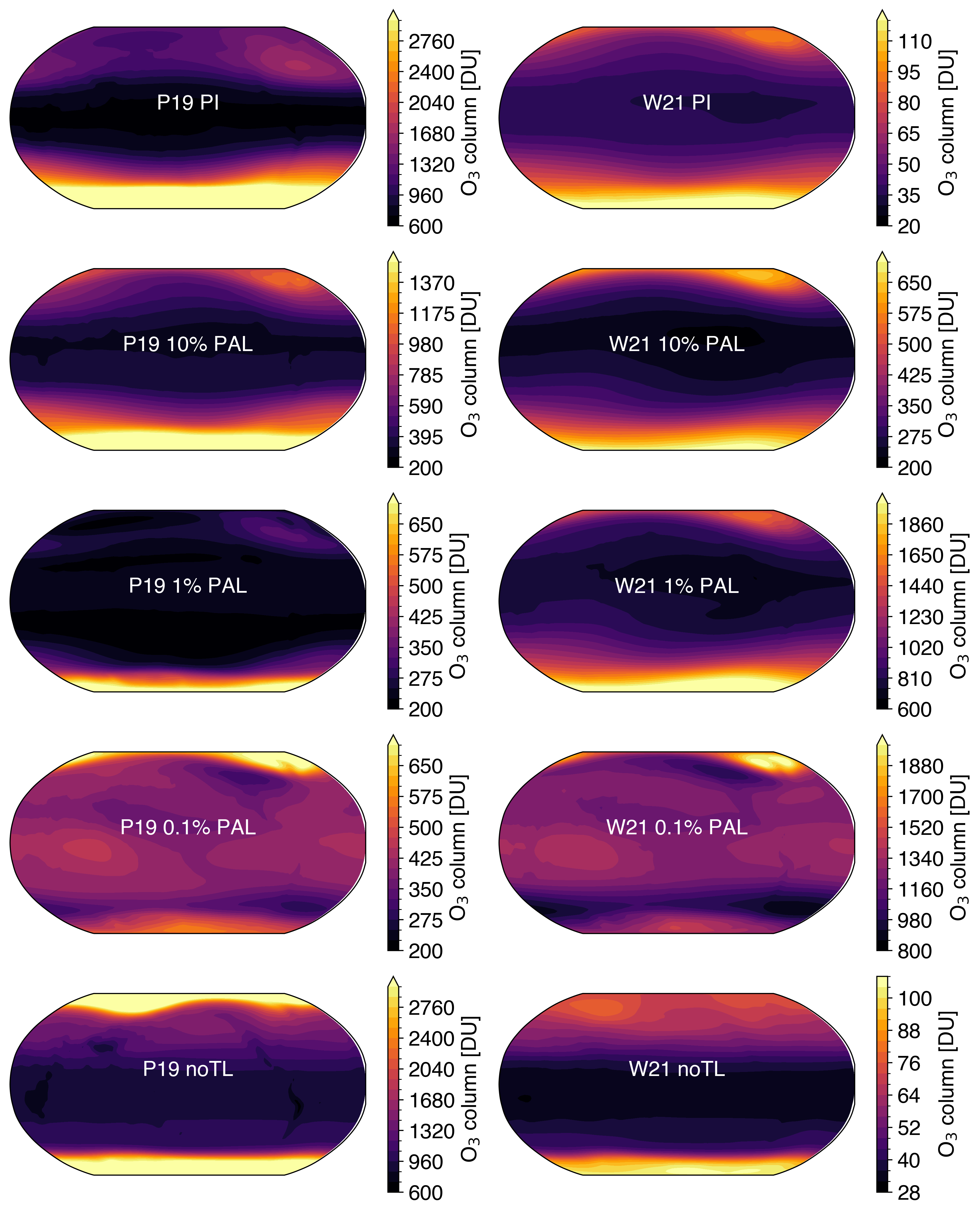}
    \caption{The \ce{O3} column in Dobson Units [DU] across the simulation's latitudinal and longitudinal grid is plotted for all the simulations used in this work (the P19 scenarios are in the left column and the W21 scenarios are in the right column). 1 DU is equal to $2.6867\times10^{20}$ molecules m\textsuperscript{-2}. In all simulations, \ce{O3} column maxima occur at high latitudes. The substellar point for the tidally locked cases is at $180^\circ$ longitude and $0^\circ$ latitude on Earth's coordinate grid (in the centre of each Robinson projection). Note that the colour bars are extended for some simulations in order to show the \ce{O3} column structure in each scenario. Each panel has a different colour bar range.} 
    \label{O3 column figure}
\end{figure*}

\begin{figure*}[t!]
	\centering
	\includegraphics[width=1\textwidth]{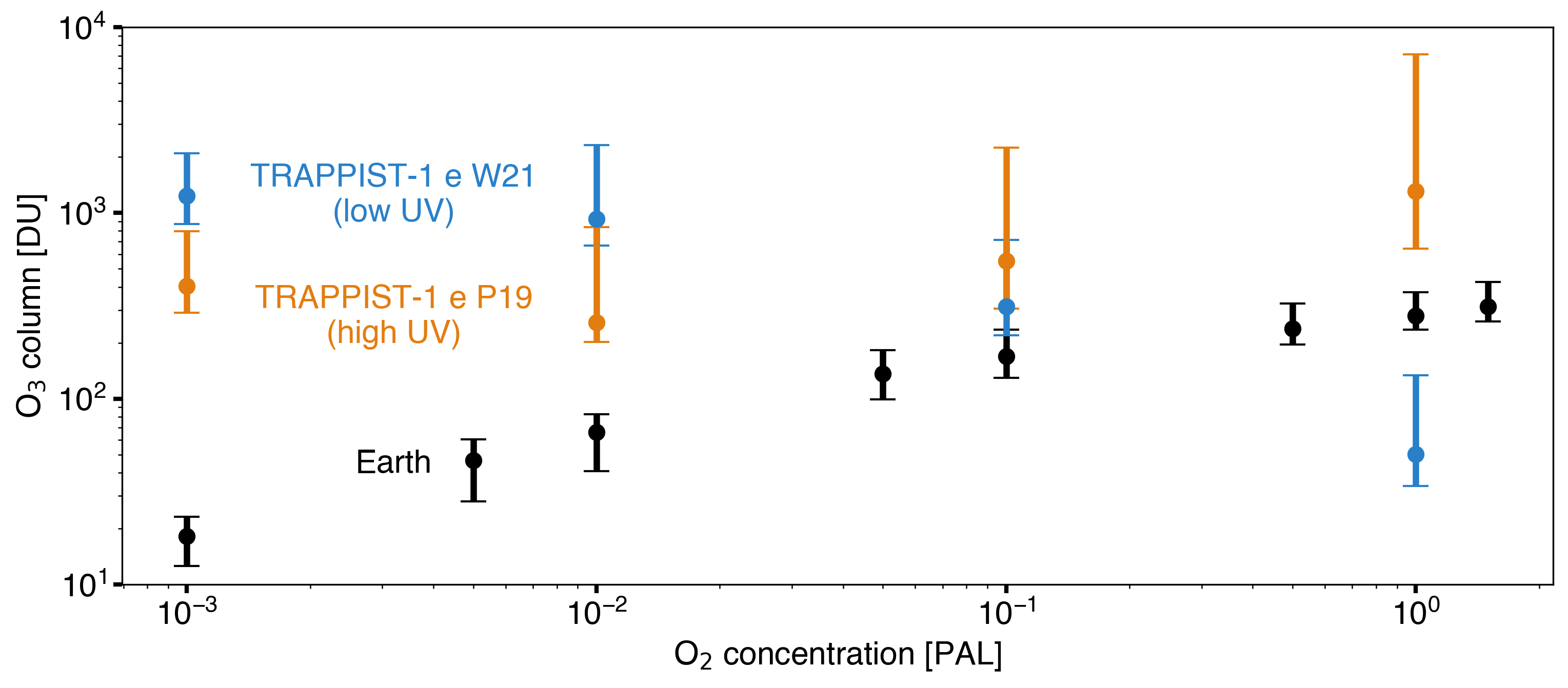}
    \caption{The oxygen (\ce{O2}) concentration is shown against the ozone (\ce{O3}) column in each simulation (orange for P19 scenarios and blue for W21 scenarios). The `error bar' lines show the range between the minimum and maximum \ce{O3} columns in each simulation. The circular dots show the mean \ce{O3} columns. The \ce{O3} columns simulated by WACCM6 for Earth, from \cite{2022RSOS....911165C}, are shown in black for comparison.}
    \label{O2 vs O3 column figure}
\end{figure*}

Fig.~\ref{O3 column figure} displays the longitudinal and latitudinal variation of the \ce{O3} column for the simulations averaged over the last ten years of each scenario. \ce{O3} is inhomogeneously distributed horizontally, which highlights the importance of using 3D models. The P19 PI simulation has a global mean \ce{O3} column of 1310 DU (approximately 4.4 times Earth's global mean value of $\approx 300$ DU, where 1 DU = $2.687\times 10^{20}$ molecules m$^{-2}$), whilst the W21 PI simulation predicts a global mean \ce{O3} column of 50 DU. Both of the \ce{O3} column maxima, at 7152 DU and 134 DU for the P19 PI and W21 PI simulations, respectively, occur near the southern pole.

Fig.~\ref{O2 vs O3 column figure} shows the total range of \ce{O3} columns produced across the atmospheric latitude-longitude grid against the \ce{O2} volume mixing ratio. The \ce{O3} column depends on the incident UV radiation, \ce{O2} concentration, atmospheric pressure, temperature, and transport, as well as the rates of destruction. In the W21 PI case, the UV radiation is absorbed high in the atmosphere where the pressure is low, so little \ce{O3} is produced and the O\textsubscript{x} (\ce{O + O3}) is primarily in O. As \ce{O2} is decreased, the \ce{O3} column increases. In contrast, in the P19 simulations, which have larger incoming fluxes of UV, as \ce{O2} increases above 1\% PAL, the \ce{O3} column increases, similar to the relationship on Earth. Both of these data sets are rather different when compared to previous results that have simulated the atmospheres of planets around late M dwarf stars  \citep[e.g.,][although note that the total instellation in those simulations were set to Earth's modern value]{2018ApJ...854...19R, 2022A&A...665A.156K}. Whilst there are likely many interacting parameters which will cause dissimilar \ce{O3} column predictions, the very large discrepancies with 1D models could be due to 3D transport. This seems plausible because \cite{2023arXiv230603004B} reported a stratospheric circulation in simulations of a terrestrial Proxima Centauri b scenario, in which the winds move \ce{O3} from the day side to the night side, with similar circulation effects occurring in the WACCM6 simulations here (the large-scale dynamics of the atmosphere will be explored in future work).

\subsection{Transmission spectra}

\begin{figure*}[t!]
	\centering
	\includegraphics[width=\textwidth]{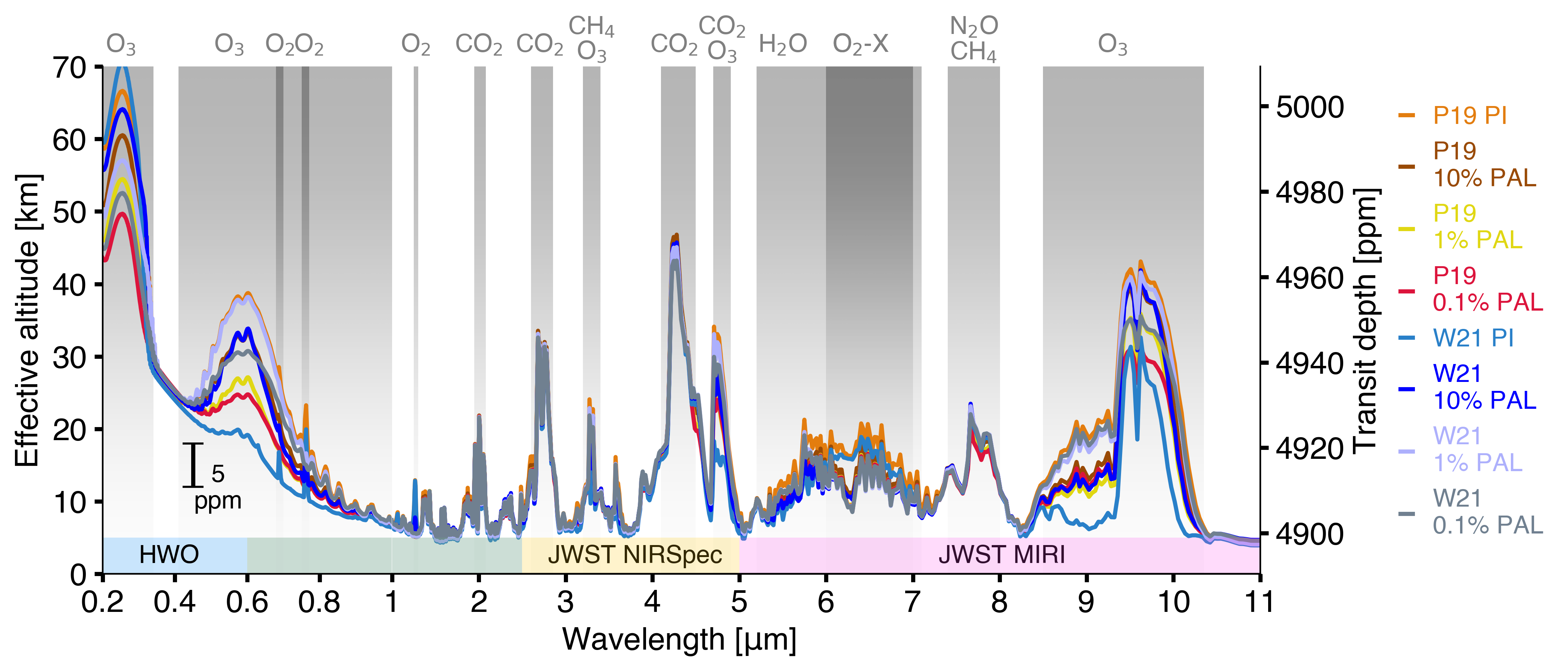}
    \caption{The transmission spectrum atmospheric effective altitude is plotted against wavelength between 0.2 \textmu m and 11 \textmu m for the P19 PI (orange), P19 10\% PAL (brown), P19 1\% PAL (yellow), and P19 0.1\% PAL (red) simulations, and the W21 PI (light blue), W21 10\% PAL (blue), W21 1\% PAL (lilac), and W21 0.1\% PAL (grey) simulations. The non-tidally locked cases are excluded for clarity, but show little differences compared to the equivalent tidally locked case. The transit depth, in terms of contrast with respect to the star, is indicated on the right vertical axis in parts per million (ppm). The spectra are binned to a spectral resolving power of $R=250$. Spectral features are indicated in grey. The wavelength range of the proposed Habitable Worlds Observatory (HWO; blue shaded region), and that of the JWST NIRSpec instrument (yellow shaded region), are shown. They overlap in the green shaded region. The wavelength range of the JWST MIRI instrument is indicated in the magenta shaded region. The wavelength spacing between 0.2 -- 1 \textmu m is changed between 1 -- 11 \textmu m in order to clearly show the UV and visible regions. The black bar represents the uncertainty that would be present on a measurement that has reached the noise floor of the instrument, where the noise floor is indicated as 5 ppm. Note this error bar is an estimate of the performance of the telescope and does not indicate a measurement.} 
    \label{Transmission spectra figure}
\end{figure*}

In Fig.~\ref{Transmission spectra figure} we show idealised transmission spectra between 0.1 -- 11 \textmu m generated using the WACCM6 simulations with PSG (excluding the non-tidally locked cases). The model date chosen for the transit is arbitrary. Because there are time-dependent fluctuations for several variables in the WACCM6 simulations (e.g. clouds, chemical mixing ratios), we could have investigated time variability in the transmission spectra. However, \citet{2022PSJ.....3..213F} showed that such variability is within the measurement uncertainties of JWST. The spectra are binned to approximate a resolving power of $R=250$ to show detail in spectral features, where $R=\lambda / \Delta \lambda$, $\lambda$ is the wavelength, and $\Delta \lambda$ is width of the wavelength bin. A 5 ppm error bar corresponding to the lowest achievable noise with JWST instruments, which may be between 5 -- 20 ppm as calculated by \citet{2019PASP..131l4502M}, \citet{2020AJ....160..231S}, \citet{2021AJ....161..115S}, and \citet{2022ApJ...928L...7R}, is indicated.

The differences in the effective altitude of \ce{O3} spectral features between the P19 PI (orange) and W21 PI (light blue) transmission spectra are $-4$ km, $+19$ km, $+15$ km, and $+17$ km for the 0.25 \textmu m, 0.6 \textmu m, 4.71 \textmu m, and 9.6 \textmu m \ce{O3} features, respectively. Despite the W21 PI simulation having an \ce{O3} column $\approx26$ times lower than the P19 PI simulation, the W21 PI UV feature (centred at 0.25 \textmu m) due to the Hartley band (0.2 -- 0.31 \textmu m) actually has the largest effective altitude between 0.2 -- 0.3 \textmu m of all the simulations. This is because the Hartley band saturates quickly and the W21 PI atmosphere has more total \ce{O3} than the P19 PI atmosphere above $\approx0.5$ hPa. Between 0.3 -- 0.35 \textmu m, the temperature dependence of the Hartley and Huggins bands reduces the effective height of the W21 PI transmission spectra due to the colder middle atmosphere in the W21 PI simulation. At 0.6 \textmu m, a significant detection of \ce{O3} with JWST in the W21 PI simulation scenario would be improbable given that the noise floor is larger than the height of the feature. Therefore, assuming that the W21 spectrum is closest to the true spectrum of TRAPPIST-1, or the case that the true stellar UV emission is weaker, a null detection of the 0.6 \textmu m \ce{O3} feature should not rule out the presence of \ce{O2} abundances at levels as high as the present-day Earth. For \ce{H2O}, the spectral features are stronger in the P19 PI transmission spectra compared to W21 PI by up to 8 km which is a result of a larger number density of \ce{H2O} in the middle atmosphere. Despite the difference in temperature and \ce{O3} number density profiles between the noTL simulation and PI simulations, the transmission spectra are remarkably similar (within $\pm$4 km effective altitude, not shown for clarity). The P19 0.1\% PAL transmission spectrum (red) produces a quantitatively similar transmission spectrum feature at 9.6 \textmu m to the W21 PI simulation (light blue), even though there is a 1000 times difference in \ce{O2} mixing ratio between the two cases. The same can be said of the following pairs at 9.6 \textmu m: W21 1\% PAL and P19 PI; W21 0.1\% PAL and P19 1\% PAL; W21 10\% PAL and P19 10\% PAL. There is a noticeable difference between the spectra at 4.71 \textmu m and 9 \textmu m, but this would require reaching the most optimistic 5 ppm noise floor in order to demonstrate that the two \ce{O2} scenarios are distinguishable when there exist uncertain stellar UV flux estimates. Also note that the effective height of the \ce{O2-X} collision-induced absorption feature at 6.4 \textmu m \citep[see][for more details]{2020NatAs...4..372F} is 7 km shallower when \ce{O2} is at 0.1\% PAL in the P19 and W21 scenarios, compared to the PI cases.

\begin{figure*}[t!]
	\centering
	\includegraphics[width=1\textwidth]{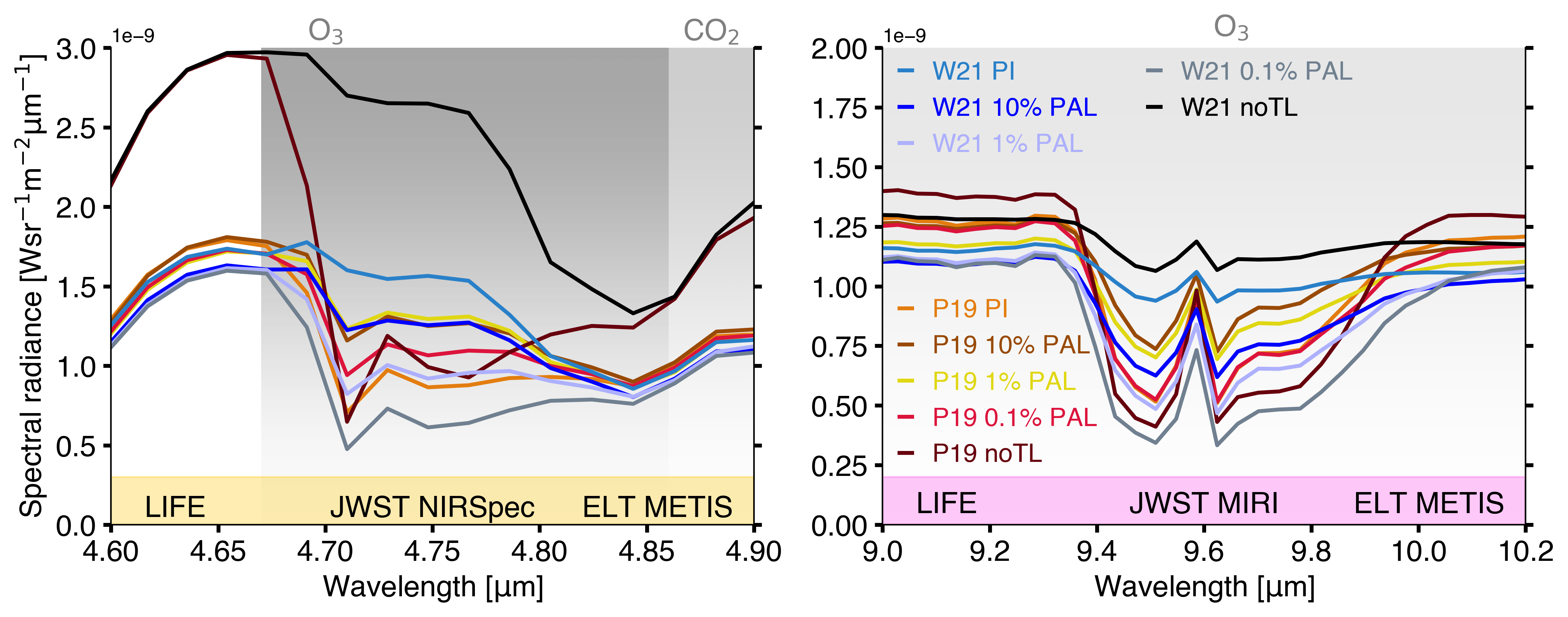}
    \caption{The left panel shows the PSG simulations of planetary spectral radiance from emission spectra focused on the 4.71 \textmu m \ce{O3} feature (which overlaps with a \ce{CO2} feature, both shown by the grey shaded regions) for the P19 PI (orange), P19 10\% PAL (brown), P19 1\% PAL (yellow), and P19 0.1\% PAL (red) simulations, and the W21 PI (light blue), W21 noTL (black), W21 10\% PAL (blue), W21 1\% PAL (lilac), and W21 0.1\% PAL (grey) simulations. The right panel shows the same for the 9.6 \textmu m \ce{O3} feature, shown by the grey shaded region. Telescopes and their instruments which may be able to probe atmospheres in the infrared include: the Extremely Large Telescope \citep[ELT;][]{2021Msngr.182...22B} METIS instrument, the Large Interferometer for Exoplanets \citep[LIFE;][]{2022A&A...664A..23K}, and the JWST \citep{2017ApJ...850..121M} NIRSpec and MIRI instruments.}
    \label{Emission/reflection spectra figure}
\end{figure*}

\subsection{Emission spectra}

In Fig.~\ref{Emission/reflection spectra figure} we show the emission spectra from each simulation for the atmospheric absorption \ce{O3} features at 4.71 \textmu m and 9.6 \textmu m at $90^\circ$ orbital phase (the maximum planet-star separation as viewed in an edge-on system with a circular orbit). The 4.71 \textmu m feature overlaps with a \ce{CO2} feature, but \ce{O3} is the dominant absorber at 4.71 \textmu m. The P19 PI simulation (orange) predicts higher \ce{O3} columns than the W21 PI case (light blue); hence, the depth of the features relative to the continuum in the W21 PI case are weaker than the P19 PI scenario by a factor of 4.2 and 4.5 at 4.71 \textmu m and 9.6 \textmu m, respectively. With respect to the P19 0.1\% PAL emission spectrum (red), the P19 PI emission spectrum (orange) has a greater depth by a factor of 1.3 and 1.1 at 4.71 \textmu m and 9.6 \textmu m, respectively, even though the P19 PI simulation has a mean \ce{O3} column which is 3.2 times higher than the P19 0.1\% PAL simulation (1310 DU versus 405 DU). The 3D effects are important here because the largest \ce{O3} columns are found at the poles, but PSG is integrating over the whole observable disk. At 4.71 \textmu m, the noTL absorption features are deeper than the tidally locked PI cases by a factor of 2.0 and 1.2 in the P19 and W21 scenarios, respectively. At 9.6 \textmu m, these values are 1.2 and 1.0. In terms of the relative depths of \ce{O3} features shown in Fig.~\ref{Emission/reflection spectra figure}, increasing depth is generally shown with increasing mean \ce{O3} columns (see Fig.~\ref{O3 column figure}).

The tidally locked simulations exhibit strong convection and high clouds around the substellar point, whereas the P19 noTL simulation has mainly low clouds with comparatively little high cloud coverage. Therefore, this results in the P19 noTL simulation (dark red) having the deepest \ce{O3} emission spectral features.

It is important to note that other orbital phases may show quantitatively different results because variability in spectral features may occur due to climate variations throughout the orbit \citep[e.g.][]{2023MNRAS.518..206C}. Additionally, longer term variations may be expected due to the possible presence of a `longitudinally asymmetric stratospheric oscillation' \citep{2022ApJ...930..152C}.

\vspace{3mm}
\section{Discussion}
\label{Discussion}

We have demonstrated that large differences in assumed stellar UV spectra can lead to different predictions for the strength of \ce{O3} spectral features. These spectral features may overlap at an assumed minimum observational uncertainty of 5 ppm, despite the fact that the \ce{O2} concentration differs by factors of up to 1000. In this section, we compare our results to previous work, consider known uncertainties, and discuss work that should be done in preparation for future exoplanet observations. 

1D photochemical modelling of M dwarf terrestrial exoplanet atmospheres has demonstrated that \ce{CH4} and \ce{N2O} could have greater abundances in the middle atmosphere compared to the modern Earth's atmosphere \citep[e.g.][]{2005AsBio...5..706S,2019A&A...624A..49W}, which we also find in our WACCM6 simulations of TRAPPIST-1e (see Fig.~\ref{Chemistry figure}). \citet{2022ApJ...927...90T} showed changes of over two orders of magnitude in the middle atmosphere \ce{O3} mixing ratios when modelling a modern Earth-like exoplanet that receives $1\ S_\oplus$ of irradiation around GJ 176 (an M2.5V star), but with various UV irradiation scenarios. They derived transmission spectra predictions from their atmospheric simulations and found the maximum transit depth differences to be $<2$ ppm, which is below the noise floor (5 ppm or greater) for JWST. This is in contrast to the W21 PI and P19 PI transmission spectra results shown here, where the strengths of estimated \ce{O3} features here are distinct at the 5 ppm level. Whilst different atmospheric modelling methods are used in \citet{2022ApJ...927...90T} compared with this work, the difference in observational significance is primarily due to the size of the stars modelled: the radius of GJ 176 is $0.45\ R_\odot$, and the radius of TRAPPIST-1 is $0.1192\ R_\odot$. \citet{2022ApJ...927...90T} also demonstrated that hazy Archean Earth atmospheres were more sensitive to changes in the incoming UV spectra compared to the modern Earth's atmosphere, which warrants future investigations for how uncertainties in the UV spectrum of the host star affect hazy atmospheres in 3D models.

In terms of 3D modelling, the THAI series \citep{2020GMD....13..707F,2022PSJ.....3..212S, 2022PSJ.....3..211T} has investigated the climate of TRAPPIST-1e using four different 3D GCMs, assuming either an \ce{N2} or \ce{CO2} dominated atmosphere and not including interactive chemistry, where the composition evolves depending on chemical and photochemical reactions. The surface temperatures in the WACCM6 simulations are similar although slightly lower (219 -- 231 K global mean compared to 230 -- 240 K in the THAI simulations. See appendix \ref{Surface temperatures appendix} and Fig.~\ref{Surface temperatures figure} for more details), which may be due to differences in assumptions regarding the surface (including the distribution of the continents and the fact that an interactive ocean is used here, in contrast to a slab ocean with no meridional heat transport) or the composition of the atmosphere. With a previous version of WACCM (CESM1), \citet{2019ApJ...886...16C} investigated a planet with a 43.87 day orbital period around a star with an effective temperature of 4000 K and an insolation of $1.9\ S_\oplus$, as opposed to the $0.66\ S_\oplus$ used here, and assessed the impact of uncertain host-star UV flux on the atmosphere. \citet{2019ApJ...886...16C} showed that two different spectra (representing a quiescent and an active M dwarf star) impacted the middle atmospheric concentrations of \ce{O3}, \ce{OH}, \ce{N2O}, \ce{CH4}, and \ce{H2O}. They calculated transmission spectra for the two simulated atmospheres, finding that the only observable difference was for the \ce{O3} feature at 9.6 \textmu m (although the UV \ce{O3} feature is not shown in their figure 11).  On the other hand, the transmission spectra simulations shown here in Fig.~\ref{Transmission spectra figure} display noticeable spectral differences for \ce{O3} at 0.25, 0.6, 4.7, 9.0 and 9.6 \textmu m, as well as for \ce{H2O} between 5 -- 6 \textmu m. The differences in predicted observations between our work and that of \citet{2019ApJ...886...16C} likely arise due to the differences in exoplanetary system setup, the different stellar spectra, and the calculated lower \ce{O3} columns from \citet{2019ApJ...886...16C}, compared to the simulated atmospheres here. 43.87 days is in the `slow rotator' regime \citep[for the definition of tidally locked rotation regimes see][]{2018ApJ...852...67H}, and 6.1 days for TRAPPIST-1e can correspond to either the `Rhines rotator' or `fast rotator' regime \citep{2022PSJ.....3..214S}. Thus, our results, alongside those from \citet{2019ApJ...886...16C}, demonstrate that 3D modelling results are sensitive to the choice of the assumed stellar UV spectra for potentially habitable tidally locked exoplanets across early and late M dwarf stars and different rotation periods. Future work should also investigate the influence of orbital perturbations away from a synchronous 1:1 spin-orbit resonance \citep[e.g.][]{2023arXiv230211561C} on composition.

The anisotropy of chemical molecules will affect the emission spectra and photometry of terrestrial exoplanets, by modulating when the exoplanet is brightest and dimmest at particular wavelengths \citep{2011A&A...532A...1S, 2018ApJ...868L...6C}. The phase curve photometry amplitude, and where the maximum brightness occurs during the orbit, depends on the particular rotation regime that the exoplanet exists within \citep{2018ApJ...852...67H}. Thermal emission features are influenced by molecular abundance, the atmospheric temperature, and the temperature difference between the emitting and absorbing region (e.g., for Earth, the infrared emission emanates from the troposphere, whilst the absorbing region is the \ce{O3} layer in the stratosphere). This temperature difference is larger for M dwarf exoplanets which exhibit a less pronounced stratospheric temperature inversion due to lower incident UV emission for the same total instellation.

Note that detecting \ce{O3} will be difficult with JWST within the nominal 5 year mission lifetime (although JWST is expected to continue science operations for at least 10 years), even for a modern Earth scenario \citep{2021MNRAS.505.3562L}, and \citet{2019ApJ...887..194F} found that gases other than \ce{CO2} may require hundreds or thousands of transits to be detectable. Simulations of high-resolution observations with the extremely large class of telescopes indicate that \ce{O2} at 0.76 \textmu m may be detectable in the case of TRAPPIST-1e within $\sim 100$ transits \citep{2013ApJ...764..182S,2014ApJ...781...54R,2019ApJ...871L...7S}.

The derived Mega-MUSCLES spectrum of TRAPPIST-1 \citep[W21; ][]{2021ApJ...911...18W} is constrained by more observations than the P19 spectrum, but both spectra have significant flux uncertainties. Whilst neither spectrum used in this study is likely to wholly represent the true stellar irradiation environment of TRAPPIST-1e, there are at least observational constraints on the `ground truth' of its parent star's spectrum. For many planetary systems, there will only be estimates from stellar models, and this will cause significant problems for predicting the photochemical environment of potentially habitable exoplanets. Furthermore, in each wavelength bin, we have assumed that the flux does not vary with time. Due to M dwarf stellar activity, such an assumption is unlikely to be accurate \citep{2018ApJ...867...71L}. The \ce{O3} abundance will be perturbed by the inclusion of incident stellar flares \citep{2010AsBio..10..751S, 2019AsBio..19...64T, 2021NatAs...5..298C, 2022arXiv221013257R} which we have not investigated here. Based on previous results, it seems that stellar flares will exacerbate the interpretation of observed spectra, so future work on incoming UV uncertainties could evaluate the additional impact of stellar flares. The present modelling uncertainties in the \ce{O2}-\ce{O3} non-linear relationship arising from differences in predictions between 1D and 3D models \citep{2022RSOS....911165C, 2022A&A...665A.156K, 2022CliPa..18.2421Y, 2023RSOS...1030056J} will compound this issue. In addition to previous work, our simulations, which focus on the specific target of TRAPPIST-1e, further motivates the need for a dedicated next generation observatory with UV capabilities to characterise exoplanet host stars.

UV flux measurements from a telescope such as the $\sim 6$ m UV/VIS/NIR telescope (currently referred to as the Habitable Worlds Observatory) that was recommended by the Decadal Survey \citep{2021pdaa.book.....N} will aid the interpretation of observed exoplanet spectra and help to infer the  concentration of \ce{O2} and trace gases in the atmosphere without direct measurements \citep{2022A&A...665A.156K}. However, this telescope is not expected to be operational until the late 2030s at the earliest. Determining the EUV fluxes from a host star \citep[which will require a dedicated observatory;][]{2019BAAS...51c.300Y} will also provide important information about atmospheric escape, habitability, and help to examine the atmospheric history of the exoplanets in the system.

Before next generation telescopes are online, there are other clues available to characterise oxygenated terrestrial atmospheres if the interpretation of the spectral features (e.g. \ce{O3}) leaves degeneracies in the parameter space between \ce{O2} concentration, \ce{O3} concentration, UV irradiation, and \ce{O3} depleting catalytic cycles. For example, the major differences between the P19 PI and the P19 0.1\% PAL transmission spectra are between the \ce{H2O}, \ce{O2}, and \ce{O3} features. Moreover, the estimated inter-simulation trends with wavelength in transmission spectra are not mirrored in emission spectra predictions. Namely, the depth relative to the continuum in emission spectra for \ce{O3} at 4.71 \textmu m and 9.6 \textmu m contrasts with the relative strength of associated transmission spectra features between the simulations. This means that if both transmission spectra and emission spectra are acquired with adequate precision, multi-wavelength observations combined with atmospheric retrieval methods \citep{2021ExA...tmp..118Q} will be useful when delineating between possible atmospheric composition scenarios. Nevertheless, \citet{2018ApJ...856L..34B} showed that confident estimates on atmospheric composition from emission spectra observed with JWST MIRI LRS will prove difficult to achieve for temperate exoplanets, using TRAPPIST-1 f as an example. 

Finally, future modeling work should explore the impact of chemical boundary conditions on simulated transmission and emission spectra. For example, the choice of boundary conditions, such as the upward flux and abundances of species in the HO\textsubscript{x}, NO\textsubscript{x}, ClO\textsubscript{x}, and BrO\textsubscript{x} families, will not only affect the \ce{O3} distribution but also the atmospheric temperature and incoming radiation. Provided the atmospheric profile of \ce{O3} derived from atmospheric retrievals is well-constrained, by assuming minimal and maximal \ce{O3} catalytic cycle destruction, limits could be placed on \ce{O2}.

\section{Conclusions}
\label{Conclusions section}

For the first time using a 3D chemistry-climate model (WACCM6) to simulate TRAPPIST-1e (assuming an initial Earth-like composition) and including two different incoming UV spectra, we demonstrated that using a single observed \ce{O3} feature outside of the UV range to extrapolate to undetected molecules, such as \ce{O2}, will lead to degeneracies over multiple orders of magnitude in the parameter space for atmospheric composition. The UV spectrum in both of the incoming stellar spectra varies by up to a factor of $\approx 500$ for important photolysis bands, and up to $\approx 5000$ for individual wavelength bins. Whilst the atmospheric columns of many species (including \ce{O2} and \ce{CO2}) are virtually unaffected by the difference between the two spectra, for an \ce{O2} mixing ratio of 0.21, the \ce{O3} columns differ by a factor of 26 due to different \ce{O3} production rates that are sensitive to the incoming spectrum. Consequently, the associated \ce{O3} transmission spectral features differ in effective altitude by up to $19$ km, whilst the \ce{O3} features in emission spectra differ by a factor of up to 4.5 in relative depth. One implication is that a non-detection of \ce{O3} at visible wavelengths may not indicate the absence of an oxygenated atmosphere. Furthermore, tidal locking of the model results in substantially different emission spectra features which are shallower relative to the emission continuum.

Without the direct detection of \ce{O2}, additional context for determining the oxygenation state of the atmosphere can be gained from either 1) future missions that are able to better characterise the UV spectra of faint stars, or 2) sensitive direct imaging observations combined with transmission spectra observations targeting individual features.


\section*{Acknowledgments}
G.J.C. acknowledges the studentship funded by the Science and Technology Facilities Council of the United Kingdom (STFC). C.W. acknowledges financial support from the University of Leeds and from the Science and Technology Facilities Council (grant numbers ST/T000287/1 and MR/T040726/1). This work was undertaken on ARC4, part of the High Performance Computing facilities at the University of Leeds, UK.

We would like to acknowledge high-performance computing support from Cheyenne (doi:10.5065/D6RX99HX) provided by NCAR's Computational and Information Systems Laboratory, sponsored by the National Science Foundation. The CESM project is supported primarily by the National Science Foundation (NSF). This material is based upon work supported by the National Center for Atmospheric Research (NCAR), which is a major facility sponsored by the NSF under Cooperative Agreement 1852977.

This research has made use of the NASA Exoplanet Archive, which is operated by the California Institute of Technology, under contract with the National Aeronautics and Space Administration under the Exoplanet Exploration Program.


\bibliography{references}{}
\bibliographystyle{aasjournal}

\appendix

\section{Surface temperatures}
\label{Surface temperatures appendix}

\begin{figure*}[b!]
	\centering
	\includegraphics[width=1\textwidth]{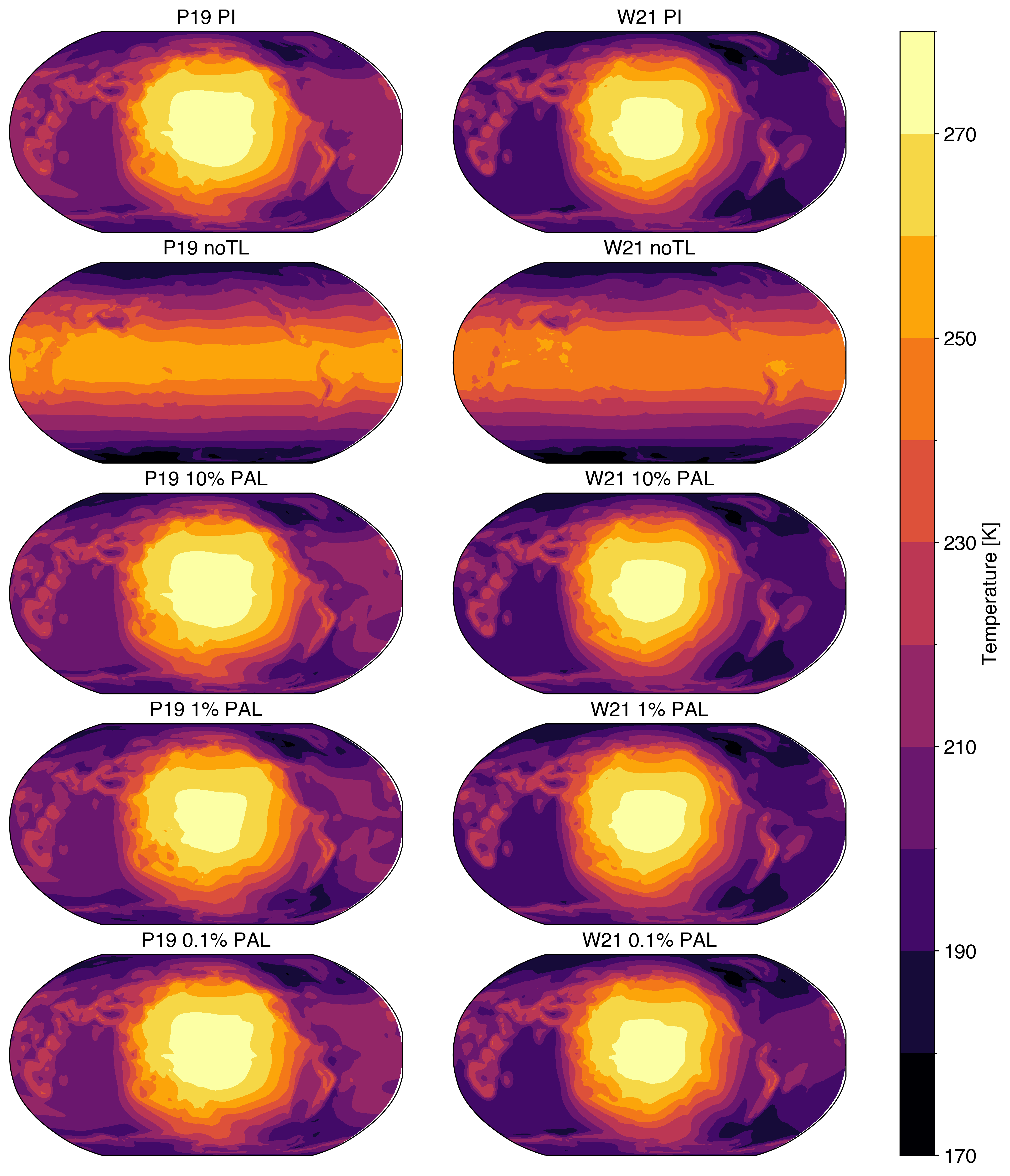}
    \caption{The surface temperatures in kelvin are shown for the P19 PI, P19 0.1\% PAL, P19 noTL, and W21 PI simulations. In the tidally locked cases, a relatively warm region exists around the substellar point, which receives constant irradiation. The substellar point for the tidally locked cases is at $180^\circ$ longitude and $0^\circ$ latitude (in the centre of each Robinson projection). In the P19 and W21 noTL cases, the substellar point moves because there is a diurnal cycle. These simulations have the coldest surface temperatures (171 and 170 K, respectively), but the smallest range in surface temperatures.} 
    \label{Surface temperatures figure}
\end{figure*}

The time-averaged surface temperatures for all the simulations (see Table \ref{Run table}
for simulation details) are shown in Fig.~\ref{Surface temperatures figure}. On time average, the global mean surface temperatures are between 219 -- 231 K, with the minima ranging between 171 -- 180 K, and the maxima ranging between 252 -- 277 K. The P19 and W21 noTL simulations are everywhere colder than 273 K, whilst all other simulations do have some areas of the surface which are warmer than 273 K (the freezing point of water under 1 bar of atmospheric pressure).

\section{Water vapour and high clouds}
\label{Water vapour column appendix}

\begin{figure*}[b!]
	\centering
	\includegraphics[width=1\textwidth]{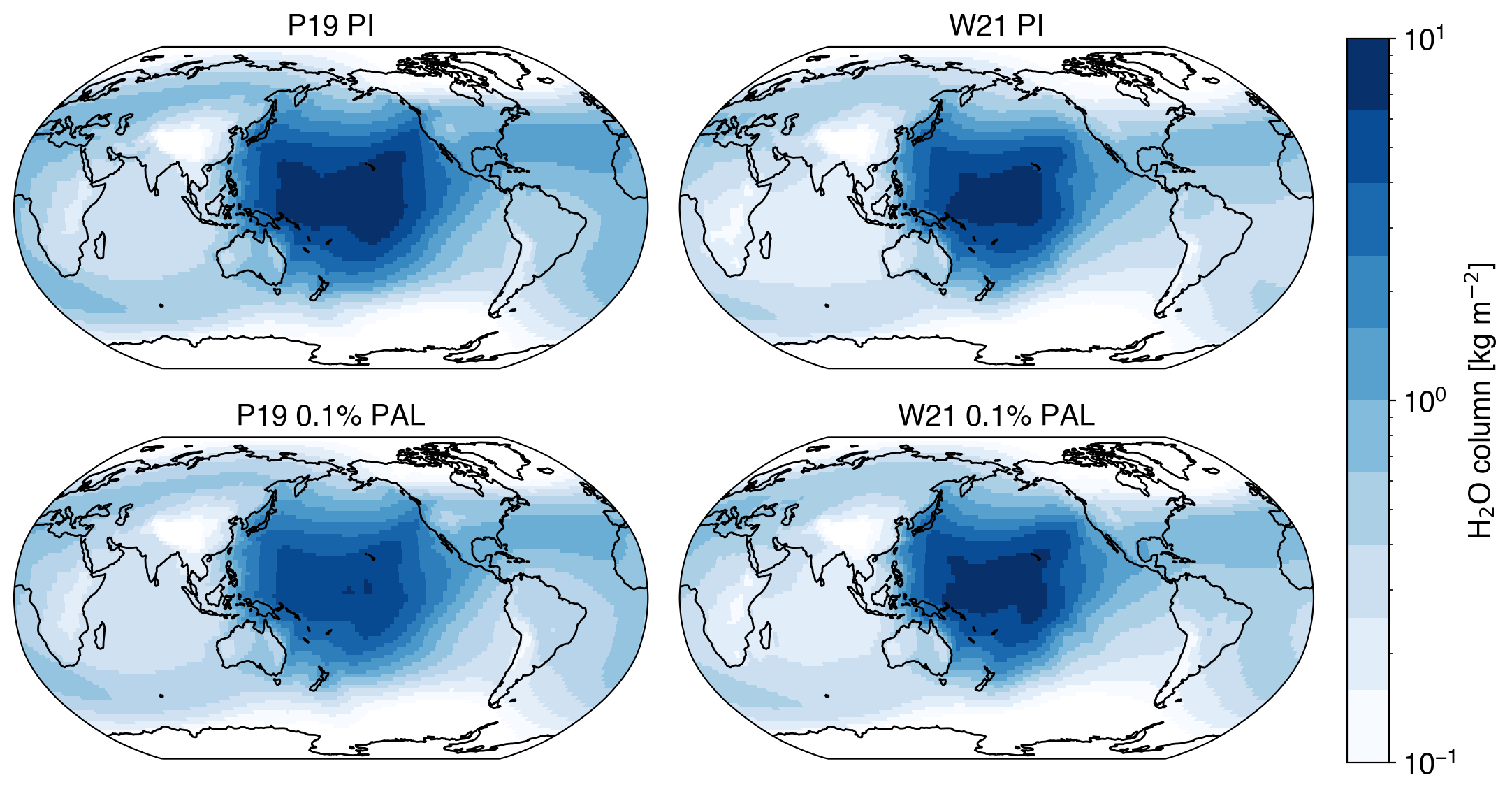}
    \caption{The \ce{H2O} column, given in terms of molecules m\textsuperscript{-2}, is shown for the PI and 0.1\% PAL simulations in the P19 (left) and W21 (right) scenarios. White shows a relatively low \ce{H2O} column, whilst blue shows relatively large amounts of \ce{H2O} column. For reference, Earth's \ce{H2O} column reaches a peak of about 50 kg m\textsuperscript{-2}.} 
    \label{H2O col figure}
\end{figure*}

Fig.~\ref{H2O col figure} shows the water vapour (\ce{H2O}) column for the PI and 0.1\% PAL simulations for both the P19 and W21 scenarios. The \ce{H2O} column is predominantly situated around and above the substellar point. This is because the incoming solar radiation heats the surface, and water vapour evaporates. The ascending air parcels carry the water vapour upwards, where the parcels cool and form clouds.

\begin{figure*}[b!]
	\centering
	\includegraphics[width=1\textwidth]{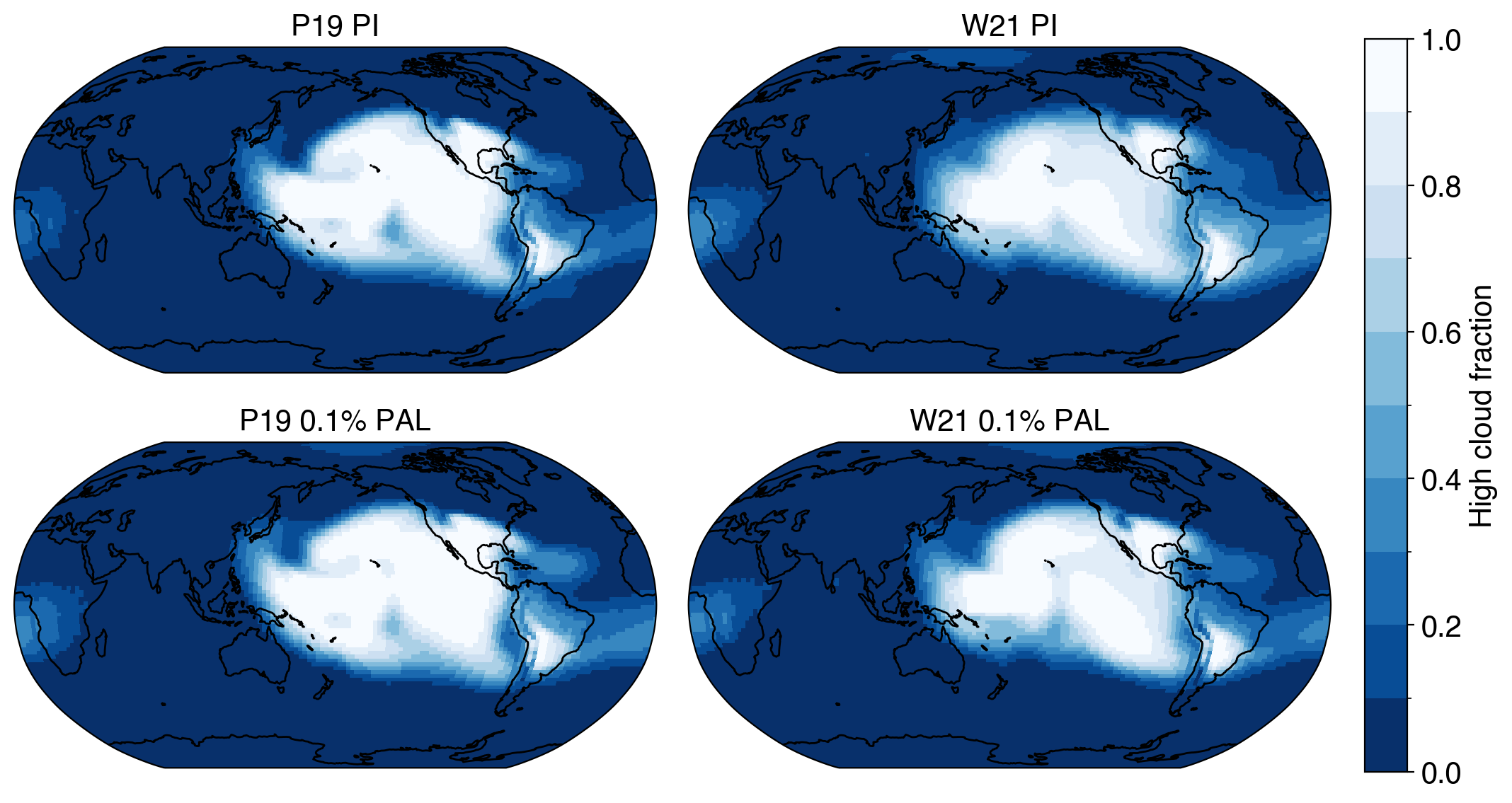}
    \caption{The high cloud fraction, where high clouds are specified as clouds below pressures of 400 hPa, is shown for the PI and 0.1\% PAL simulations in the P19 (left) and W21 (right) scenarios. Blue shows low cloud fraction, whilst white shows large amounts of cloud coverage. Clouds provide a net warming in the tidally locked scenarios.} 
    \label{Cloud fraction figure}
\end{figure*}

Fig.~\ref{Cloud fraction figure} shows the high cloud fraction for the PI and 0.1\% PAL simulations for both the P19 and W21 scenarios. High clouds are defined as clouds at pressures less than 400 hPa. Clouds, through shortwave (incoming solar radiation) and longwave (outgoing terrestrial radiation) forcing, provide a net warming of 8 -- 12 W m\textsuperscript{-2} in the tidally locked scenarios, and 5 -- 7 W m\textsuperscript{-2} in the non-tidally locked scenarios.

\end{document}